\documentclass{phostproc}

\title{He white dwarfs with large H contamination: Convective mixing or accretion?}
\author{Felipe C. Wachlin,$^{1}$ 
        G\'erard Vauclair, $^{2,3}$
        Sylvie Vauclair $^{2,3}$}

\affiliation{$^{1}$ Instituto de Astrof\'{\i}sica de La Plata (UNLP--CONICET), La Plata, Argentina \\
	     $^{2}$ Universit\'e de Toulouse, UPS--OMP, IRAP, Toulouse, France \\
	     $^{3}$ CNRS, IRAP, Toulouse, France}

\shorttitle{DB white dwarfs with large H contamination}
\shortauthors{Felipe C. Wachlin \textit{et al.}}

\abs{White dwarfs are compact objects with atmospheres containing 
mainly light elements, hydrogen or helium. Because of their surface high 
gravitational field, heavy elements diffuse downwards
 in a very short timescale compared to the evolutionary timescale, leaving
the lightest ones on the top of the envelope. This results in the main classification
of white dwarfs as hydrogen rich or helium rich. But many helium rich white dwarfs 
show also the presence of hydrogen traces in their atmosphere, whose origin 
is still unsettled. Here we study, by means of full evolutionary calculations, 
the case for a representative model of the ``He-H-Z'' white dwarfs, a sub-group of 
helium rich white dwarfs showing both heavy elements and a large amount 
of hydrogen in their atmosphere. We find it impossible to explain its 
hydrogen atmospheric
content by the convective mixing of a primordial hydrogen present in the star.
We conclude that the most likely explanation is the accretion of hydrogen
rich material, presumably water-bearing, coming from a debris disk.}

\begin{document}

\maketitle

\section{Introduction}

White dwarfs are the final evolutionary stage of most of the stars 
\citep{2010A&ARv..18..471A}. They are classified according to their 
spectroscopic appearance into two main classes. The more abundant
type ($\approx 80\%$) has H-rich atmospheres and are labeled as DA white dwarfs, 
whereas the remaining $\approx 20\%$ corresponds to those showing He-dominated 
atmospheres and are referred to as DB white dwarfs \citep{2012ApJS..199...29G,
2013ApJS..204....5K}. 

Spectroscopy cannot provide any constraint on the total fractional mass of the 
H layer and/or the He layer present in DA or DB white dwarfs. In this respect, asteroseismology 
is a powerful tool to determine these values. From asteroseismological 
studies it is founded that the hydrogen mass fraction may vary from  
$\approx$$10^{-10}$ to $\approx$$10^{-4}$ and the helium mass fraction 
is around $10^{-2}$ \citep[see for instance][]{2012MNRAS.420.1462R}.

This  picture becomes more complicated when other elements are present 
in their spectra. Particularly interesting is the case of cool white 
dwarfs showing heavy elements in their atmospheres. 
These heavy elements should indeed diffuse downwards, because of the high 
gravitational field of these compact objects, on time scales short 
compared to the evolutionary timescale. As many as 25 to 50 percent of the observed 
white dwarfs show heavy elements in their atmospheres. Some of these metallic-line white dwarfs also 
show an excess of radiation in the infrared. This is widely accepted 
as evidence of ongoing (or recent) accretion of material from debris 
disks. 

The detection of H 
traces in DB white dwarfs has also modified the initial simple picture \citep{2007A&A...470.1079V,
2011ApJ...737...28B,2015A&A...583A..86K}. The origin of this hydrogen is
still under discussion. Three different theories have been proposed
to explain it:
\begin{itemize}\itemsep=-\itemsep
 \item accretion from the interstellar medium (ISM),
 \item convective mixing of primordial H, and
 \item accretion of H-rich material from a debris disk.
\end{itemize}

Based on the largest sample of DB white dwarfs studied so far, 
\citet{2015A&A...583A..86K} concluded that the ISM accretion can be
ruled out since there is no correlation of the number of DB with H 
traces with the distance above the Galactic plane and also that 
there is no evidence that the hydrogen abundance increases with 
time in these stars.

The convective mixing theory refers to the fact that white dwarfs
develop a surface convective region deepening
as they cool down below $T_\mathrm{eff}\approx 25000$ K. 
If hydrogen is present before convection sets in, it should float
on top of helium and the white dwarf would be classified as a DA. 
On the other hand, if the H content is small and the star is 
cool enough, the H convection zone may reach the H/He transition zone, 
dredge helium up to the surface and dilute the small amount of H into
a larger (convective) region. As a result, the former DA white dwarf 
may become a DB with traces of hydrogen.
 
This seems to be a possible explanation for DBs showing small 
hydrogen traces \citep{2015A&A...583A..86K}. Here we test whether it can also explain the case of the handful of DB 
white dwarfs showing both heavy elements and a large amount of hydrogen 
in their envelope \citep[GD~16, GD~17, SDSSJ1242+5226, GD~362, PG~1225-079 
referred to as He-H-Z white dwarfs, see][]{2017MNRAS.468..971G}.  
In order to carry out this work, we perform full evolutionary simulations
(which include all relevant physical processes) to test this 
channel of formation.

%%%%%%%%%%%%%%%%%%%%%%%%%%%%%%%%%%%%%%%%%%%%%%%%%%%%%%%%%%%%%%%%%%%%%
\section{Method and results}
%%%%%%%%%%%%%%%%%%%%%%%%%%%%%%%%%%%%%%%%%%%%%%%%%%%%%%%%%%%%%%%%%%%%%

All numerical experiments were done using the {\tt LPCODE} stellar 
evolutionary code \citep{2005A&A...435..631A, 2013A&A...557A..19A}. 
The method used in this work is similar to the one described in 
\citet{2017A&A...601A..13W}.

We generated an initial model representative of the He-H-Z white dwarfs 
by adapting the 0.767 $M_\odot$ white dwarf model taken from 
\citet{2010ApJ...717..183R}, which was obtained from full 
evolutionary calculations.

The chemical composition of the accreted matter was chosen, at first, 
to mimic the bulk Earth composition \citep{2001E&PSL.185...49A}. 
At a second stage, we used a different chemical composition taken 
from \citet{2007ApJ...671..872Z}.

In order to study whether convective mixing of primordial hydrogen is able
to explain the H/He ratio observed in the He-H-Z white dwarfs, we studied a
representative model with a set of stellar parameters close to those for GD~362
\citep{2009A&A...498..517K}, thus $T_\mathrm{eff}=10540$ K, 
$M_\mathrm{wd}= 0.73 M_\odot$ and $M_\mathrm{H}=3.51\times 10^{-9} M_\odot$. We also chose an initial model with an amount of hydrogen as estimated from the observations of GD~362. 

When no accretion is taken into account, we found that helium cannot be dredged up to the outer layers. 
We then
included accretion of matter in our simulations with accretion 
rates in the range from $10^6$ to $10^{12}$ g/s, adopting the chemical 
distribution of the bulk Earth composition (not hydrogen or helium included).
Such accretion of heavy elements on
a DA white dwarf triggers fingering convection below the convective envelope which
affects the photosphere's composition 
\citep{2013A&A...557L..12D,2017A&A...601A..13W}.

\begin{figure}
	\centering
	\includegraphics[width=1.00\linewidth]{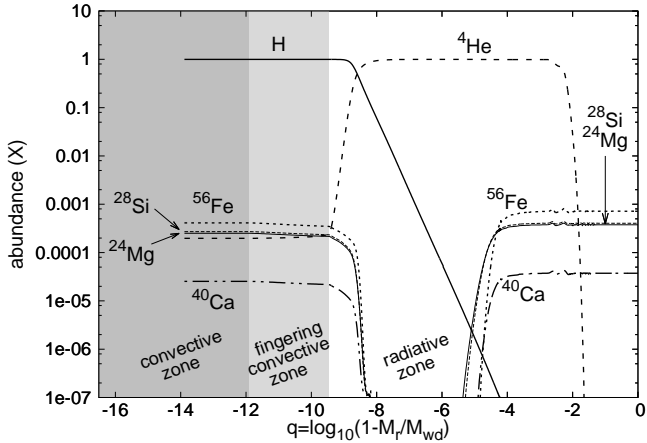}
	\caption{Final internal structure of the model subjected to an accretion
        rate of $10^{10}$ g/s.}
	\label{fig:fig_Mdot10}
\end{figure} 

Figure \ref{fig:fig_Mdot10} shows the final stellar structure of our representative
model for an accretion rate of $10^{10}$ g/s, once a stationary state has been 
achieved. Hydrogen remains the most abundant element on the surface, although 
some helium has been dredged-up. Notice that without taking into account fingering convection, no helium would have been dredged-up at all. This clearly does not 
reproduce the surface composition of the He-H-Z white dwarfs, where helium is 
the most abundant element. Increasing the accretion rate does not 
change the qualitative result.

We also explored the case of initial models with less hydrogen, such that the convection (including fingering 
convection) be able to transform the initial DA into a DB white dwarf. Therefore, we run 
a new set of simulations and found that, for this $T_\mathrm{eff}$, it is not 
possible to attain the transformation unless we reduce the amount of primordial 
hydrogen by about two orders of magnitude. Then suddenly helium becomes the most abundant element on the surface, 
much more abundant than hydrogen ($\log(n_\mathrm{He}/n_\mathrm{H})\approx 3.5$, 
where $n_\mathrm{X}$ is the density {\it by number} of element $X$). 
From \citet{2009A&A...498..517K} we know that the He-H-Z white dwarfs have a 
significantly higher H/He abundance ratio (it is 
$\log(n_\mathrm{He}/n_\mathrm{H})\approx 1.14$ for GD~362). Thus our experiment
overestimates the amount of helium dredged-up from the inner layers. Other
simulations, increasing the amount of primordial hydrogen gave similar 
results. None of them was able to better reproduce the surface relation
between hydrogen and helium. Interestingly, the experiments show
that there is no smooth transition between the two kinds of results: 
either hydrogen was dominant, or helium in an overwhelming proportion.

From these results we conclude that the theory of convective mixing of 
primordial hydrogen cannot explain the case of the He-H-Z white dwarfs. 
Since accretion from the ISM has been ruled out by 
observational arguments, this leaves us with the remaining theory that
sustains that accretion of hydrogen rich material from a debris disk 
must be responsible for the hydrogen traces observed in those stars.

%%%%%%%%%%%%%%%%%%%%%%%%%%%%%%%%%%%%%%%%%%%%%%%%%%%%%%%%%%%%%%%%%%%%%
\section{Conclusions}
%%%%%%%%%%%%%%%%%%%%%%%%%%%%%%%%%%%%%%%%%%%%%%%%%%%%%%%%%%%%%%%%%%%%%
We studied the theory of convective mixing of primordial hydrogen 
as the explanation of the He-H-Z white dwarfs, a sub-type of DB white 
dwarf which show an exceptionally high amount of hydrogen in their 
atmosphere. By performing numerical simulations we found that, for 
a typical hydrogen content estimated for such star (it is 
$M_\mathrm{H}=3.51\times 10^{-9} M_\odot$ for instance in GD~362), 
it is not possible to produce the transformation from DA to DB that 
should take place if convective mixing would be an efficient mechanism to 
dredge-up helium from the interior. By lowering the amount of hydrogen
we increase the chances to dredge-up helium, since the H/He transition zone
becomes closer to the surface. We found that it is necessary to reduce 
the amount of hydrogen by approximately two orders of magnitude in order to 
achieve the transformation of the atmosphere from hydrogen rich to 
helium rich. Unfortunately it is not possible to reproduce the
correct proportion of H/He when the transformation is successful.

We conclude from the present work that the He-H-Z white dwarfs 
represent particularly interesting objects because their existence 
seems to require the accretion of hydrogen rich material from the surroundings. 
This hydrogen rich material may come from different compounds like 
ammonia (NH$_3$) or methane (CH$_4$), but also from water-bearing 
planetesimals, as suggested earlier by \citet{2013Sci...342..218F}, 
\citet{2015MNRAS.450.2083R} and \citet{2017MNRAS.468..971G}. 
This last possibility suggest a very appealing way of exploring 
the potential of planetesimals to provide water in connection with the 
question of exo-planets habitability. These results will be presented in more details in a forthcoming paper.

\section*{Acknowledgments}
FCW gratefully acknowledges the help of the LOC.

\bibliographystyle{phostproc}
\bibliography{phost-paper.bib}

\begin{thebibliography}{18}
\providecommand{\natexlab}[1]{#1}

\bibitem[\protect\astroncite{{All{\`e}gre}
  \emph{et~al.}}{2001}]{2001E&PSL.185...49A}
{All{\`e}gre}, C., {Manh{\`e}s}, G., \& {Lewin}, {\'E}. 2001, Earth and
  Planetary Science Letters, 185, 49.

\bibitem[\protect\astroncite{{Althaus}
  \emph{et~al.}}{2010}]{2010A&ARv..18..471A}
{Althaus}, L.~G., {C{\'o}rsico}, A.~H., {Isern}, J., \& {Garc{\'{\i}}a-Berro},
  E. 2010, \aapr, 18, 471.

\bibitem[\protect\astroncite{{Althaus}
  \emph{et~al.}}{2013}]{2013A&A...557A..19A}
{Althaus}, L.~G., {Miller Bertolami}, M.~M., \& {C{\'o}rsico}, A.~H. 2013,
  \aap, 557, A19.

\bibitem[\protect\astroncite{{Althaus}
  \emph{et~al.}}{2005}]{2005A&A...435..631A}
{Althaus}, L.~G., {Serenelli}, A.~M., {Panei}, J.~A., {C{\'o}rsico}, A.~H.,
  {Garc{\'{\i}}a-Berro}, E., \emph{et~al.} 2005, \aap, 435, 631.

\bibitem[\protect\astroncite{{Bergeron}
  \emph{et~al.}}{2011}]{2011ApJ...737...28B}
{Bergeron}, P., {Wesemael}, F., {Dufour}, P., {Beauchamp}, A., {Hunter}, C.,
  \emph{et~al.} 2011, \apj, 737, 28.

\bibitem[\protect\astroncite{{Deal} \emph{et~al.}}{2013}]{2013A&A...557L..12D}
{Deal}, M., {Deheuvels}, S., {Vauclair}, G., {Vauclair}, S., \& {Wachlin},
  F.~C. 2013, \aap, 557, L12.

\bibitem[\protect\astroncite{{Farihi}
  \emph{et~al.}}{2013}]{2013Sci...342..218F}
{Farihi}, J., {G{\"a}nsicke}, B.~T., \& {Koester}, D. 2013, Science, 342, 218.

\bibitem[\protect\astroncite{{Gentile Fusillo}
  \emph{et~al.}}{2017}]{2017MNRAS.468..971G}
{Gentile Fusillo}, N.~P., {G{\"a}nsicke}, B.~T., {Farihi}, J., {Koester}, D.,
  {Schreiber}, M.~R., \emph{et~al.} 2017, \mnras, 468, 971.

\bibitem[\protect\astroncite{{Giammichele}
  \emph{et~al.}}{2012}]{2012ApJS..199...29G}
{Giammichele}, N., {Bergeron}, P., \& {Dufour}, P. 2012, \apjs, 199, 29.

\bibitem[\protect\astroncite{{Kleinman}
  \emph{et~al.}}{2013}]{2013ApJS..204....5K}
{Kleinman}, S.~J., {Kepler}, S.~O., {Koester}, D., {Pelisoli}, I., {Pe{\c
  c}anha}, V., \emph{et~al.} 2013, \apjs, 204, 5.

\bibitem[\protect\astroncite{{Koester}}{2009}]{2009A&A...498..517K}
{Koester}, D. 2009, \aap, 498, 517.

\bibitem[\protect\astroncite{{Koester} \& {Kepler}}{2015}]{2015A&A...583A..86K}
{Koester}, D. \& {Kepler}, S.~O. 2015, \aap, 583, A86.

\bibitem[\protect\astroncite{{Raddi} \emph{et~al.}}{2015}]{2015MNRAS.450.2083R}
{Raddi}, R., {G{\"a}nsicke}, B.~T., {Koester}, D., {Farihi}, J., {Hermes},
  J.~J., \emph{et~al.} 2015, \mnras, 450, 2083.

\bibitem[\protect\astroncite{{Renedo}
  \emph{et~al.}}{2010}]{2010ApJ...717..183R}
{Renedo}, I., {Althaus}, L.~G., {Miller Bertolami}, M.~M., {Romero}, A.~D.,
  {C{\'o}rsico}, A.~H., \emph{et~al.} 2010, \apj, 717, 183.

\bibitem[\protect\astroncite{{Romero}
  \emph{et~al.}}{2012}]{2012MNRAS.420.1462R}
{Romero}, A.~D., {C{\'o}rsico}, A.~H., {Althaus}, L.~G., {Kepler}, S.~O.,
  {Castanheira}, B.~G., \emph{et~al.} 2012, \mnras, 420, 1462.

\bibitem[\protect\astroncite{{Voss} \emph{et~al.}}{2007}]{2007A&A...470.1079V}
{Voss}, B., {Koester}, D., {Napiwotzki}, R., {Christlieb}, N., \& {Reimers}, D.
  2007, \aap, 470, 1079.

\bibitem[\protect\astroncite{{Wachlin}
  \emph{et~al.}}{2017}]{2017A&A...601A..13W}
{Wachlin}, F.~C., {Vauclair}, G., {Vauclair}, S., \& {Althaus}, L.~G. 2017,
  \aap, 601, A13.

\bibitem[\protect\astroncite{{Zuckerman}
  \emph{et~al.}}{2007}]{2007ApJ...671..872Z}
{Zuckerman}, B., {Koester}, D., {Melis}, C., {Hansen}, B.~M., \& {Jura}, M.
  2007, \apj, 671, 872.

\end{thebibliography}

\end{document}